\documentclass[aps,twocolumn,amsmath,amssymb]{revtex4}
\usepackage{psfig}
\usepackage{bm}% bold math

\renewcommand{\b}{\bf}

\begin{document}
\title{Potential energy landscape-based extended van der Waals equation}
\author{T. Keyes and J. Chowdhary}
\affiliation{Department of Chemistry, Boston University, Boston, MA 02215}
\date{\today}

\begin{abstract}
The inherent structures ({\it IS}) are the local minima of the potential energy surface or landscape, $U({\bf r})$, of an {\it N} atom system. Stillinger has given an exact {\it IS} formulation of thermodynamics. Here the implications for the equation of state are investigated. It is shown that the van der Waals ({\it vdW}) equation, with density-dependent $a$ and $b$ coefficients, holds on the high-temperature plateau of the averaged {\it IS} energy. However, an additional ``landscape'' contribution to the pressure is found at lower $T$.  The resulting extended {\it vdW} equation, unlike the original, is capable of yielding a water-like density anomaly,  flat isotherms in the coexistence region {\it vs} {\it vdW} loops, and several other desirable features. The plateau energy, the width of the distribution of {\it IS}, and the ``top of the landscape'' temperature are simulated over a broad reduced density range, $2.0 \ge \rho \ge 0.20$, in the Lennard-Jones fluid. Fits to the data yield an explicit  equation of state, which is argued to be useful at high density; it nevertheless reproduces the known values of $a$ and $b$ at the critical point.
\end{abstract}
\maketitle
\section{Introduction \label{intro}}
Classical thermodynamics and dynamics are ultimately governed by the potential energy surface or landscape, denoted {\it PEL}, defined at constant volume in the $3N$ dimensional configuration space of all the atomic coordinates of an atomic or molecular system. The canonical partition function is determined by the integral of the Boltzmann factor over the configuration space. Stillinger and Weber \cite{swIS} proposed a novel approach to this longstanding theoretical challenge. The space is partitioned into the basins of attraction of the local minima, named \cite{swIS} inherent structures ({\it IS}), and the integral becomes a sum over basins. Knowledge of distribution in energy of distinguishable {\it IS} and of the ``vibrational free energy'' of the basins, $A_{v}(U_{is},T,V)$ then allows \cite{swIS,pdb,pdbeos} evaluation of the partition function, the Helmholtz free energy $A(T,V)$,  and all of thermodynamics. Of course, obtaining these {\it PEL} properties is difficult, and detailed implementation of the {\it IS} formalism is in its infancy.

Intensive quantities are defined \cite{swIS,pdb} as $\phi$=$U_{is}/N$, $a_{v}$=$A_{v}/N$. The {\it IS} distribution is $\Omega(\phi)=Cexp(N\sigma(\phi))$, where $C$ has dimensions of inverse energy. Then,
\begin{equation}
A(N\phi,T,V)/N=\phi+a_{v}(\phi,T,V) - k_{B}T\sigma(\phi,V),
\label{A1}
\end{equation}
and a plausible, but not unique \cite{jctk3}, expression for the configuration entropy/particle is determined by $\sigma$, $S_{c}(N\phi,V)/N= k_{B}\sigma(\phi,V)\equiv s_{c}(\phi,V)$. In the limit of large {\it N} the {\it IS} energy which minimizes $A$, the thermodynamic average denoted $\phi(T,V)$, is overwhelmingly dominant. The thermodynamic $A$ is obtained by replacing $\phi$ with $\phi(T,V)$ in Eq~\ref{A1}.

The equation of state \cite{pdb,emeos} is determined by the relation, $P=-(\partial A/\partial V)_{T}=+\rho^{2}(\partial (A/N)/\partial \rho)_{T}$, where $\rho$ is the number density; we will use density instead of volume in the following. Given $ \phi(T,\rho)$, $\sigma(\phi,\rho)$, and $a_{v}(\phi,T,\rho)$, one can simply evaluate Eq~\ref{A2} and differentiate. Alternatively, note that the density  derivative of $A$ acts both on the explicit $\rho$-dependence of $\sigma$ and $a_{v}$, and on that implicit in $\phi(T,\rho) $. Using $(\partial A/\partial \phi)$=0 at equilibrium it may be seen that the implicit contributions sum to zero and, 
\begin{eqnarray}
P=\rho^{2}((\frac{\partial a_{v}}{\partial \rho})( \phi(T,\rho),T,\rho) \nonumber \hspace{.6in}  \\ +(\frac{\partial \sigma}{\partial \rho})( \phi(T,\rho),T,\rho),\hspace{.4in}
\label{P1}
\end{eqnarray}
including the explicit $\rho$-dependence only.

Little is currently known about these quantities. A Gaussian approximation,
\begin{equation}
 \sigma(\phi,\rho)=\alpha-(\phi-\phi_{0}(\rho))^{2}/2\delta^{2}(\rho),
 \end{equation}
where $\phi_{0}$ is the band center and $\delta^{2}$ is the standard deviation squared/atom, is reasonable for fluids if one believes \cite{derrida,bhup,bw,rem} that the {\it IS} are built up from weakly interacting local regions. The total number of {\it IS} is \cite{swIS} $\propto exp(\alpha N)$. 

A harmonic approximation to $a_{v}$ is natural at higher densities, and could be quite good at low $T$,
\begin{eqnarray}
a_{v}(\phi,T,\rho)=\frac{k_{B}T}{N}\langle \sum_{i=1}^{3N}ln(\beta \hbar \omega_{i}) \rangle (\phi)\nonumber \hspace{.6in} \\= k_{B}T[3 ln(\beta \hbar \omega_{0})+ \frac{1}{N} \langle \sum_{i=1}^{3N}ln(\omega_{i}/\omega_{0}) \rangle (\phi)],
\label{harm}
\end{eqnarray}
where $\omega_{i}$ is the $i'th$ normal mode frequency, $\omega_{0}$ is the frequency unit, and the averages are over representative {\it IS} with energy $\phi$. Sastry \cite{srinat} found that the frequency sum has a linear $\phi$-dependence, 
\begin{equation}
\frac{1}{N} \langle \sum_{i=1}^{3N}ln(\omega_{i}/\omega_{0}) \rangle (\phi)=f^{0}(\rho)+f^{1}(\rho) \phi,
\end{equation}
 for some liquid states of the Lennard-Jones (LJ) mixture. 

La Nave et al. \cite{emeos} reached the same conclusion for OTP, and obtained and tested the Gaussian - linear harmonic (denoted GLH) equation of state. With these approximations,
\begin{equation}
 \phi(T,\rho)=(\phi_{0}(\rho)-f^{1}(\rho)\delta^{2}(\rho))-\frac{\delta^{2}(\rho)}{k_{B}T}.
 \label{phiT1}
 \end{equation}
The pressure has the form \cite{emeos}
\begin{equation}
 P(T,\rho)=TP_{T}(\rho) + P_{const}(\rho) + T^{-1}P_{1/T}(\rho),
 \end{equation}
 where
\begin{eqnarray}
P_{T}=-\rho^{2} \frac{\partial}{\partial \rho}(s_{c,\infty}-k_{B}(f^{0}+f^{1} \phi_{\infty})) \label{PT}\\
P_{const}=\rho^{2}(\partial \phi_{\infty}/\partial \rho) \label{Pc} \\
P_{1/T}=-\rho^{2}\frac{\partial}{\partial \rho}(\frac{\delta^{2}}{2k_{B}}). \label{P1/T}
\end{eqnarray}
The {\it IS} energy reaches a high-$T$ plateau,  $\phi_{\infty}=\phi_{0}-f^{1}\delta^{2}$, and, correspondingly, $s_{c,\infty}=k_{B}\sigma(\phi_{\infty},\rho)$.

The \cite{emeos} ``potential energy landscape equation of state'', evaluated via computer simulation, was shown to accurately reproduce the true pressure for a range of liquid-state $\rho$ and $T$ in OTP.  Here we attempt to reach some more general conclusions. We identify the origin of the van der Waals equation in the {\it IS} formalism, and obtain expressions for density-dependent ``$a$'' and ``$b$''  coefficients. There is an additional contribution, $P_{1/T}$ in the GLH approximation, which we call the ``landscape pressure''. The resulting extended van der Waals equation has the possibility of yielding a water-like density anomaly (already pointed out by Sciortino et al. \cite{anomaly}), flat isotherms in the coexistence region, as opposed to {\it vdW} loops, a positive derivative $(\partial U/\partial \rho)_{T}$ at high density, and a critical anomaly in $C_{V}$.  We make use of the GLH approximation but also derive results independent of any assumptions about the vibrational free energy.  Relevant {\it PEL}  quantities are determined by computer simulations on the single-component LJ fluid for a wide range of densities including the coexistence region. Fits to the data lead to to an analytic equation of state.

\section{The extended van der Waals equation of state}
In the {\it vdW} equation the pressure,
\begin{equation}
 P(T,\rho)=\frac{k_{B}T\rho}{1-b\rho} -a\rho^{2},
 \end{equation}
is a sum of a $T$-independent term and a linear term. Here $a$ and $b$ are the well known coefficients expressing the reduction of $P$ from the ideal gas value by the attractive forces and the increase due to repulsions, respectively. 

\smallskip
\noindent {\it 1. The Gaussian linear harmonic (GLH) approximation}
\smallskip

Identification of the attractive and repulsive-ideal gas contributions with $P_{const}$ and $P_{T}$ is obvious, and holds up upon further consideration. The standard expression is $a=-(\partial (U/N)/\partial \rho)_{T}$, where $U$ is the total energy. In the GLH approximation, $U=U_{is}+\frac{3}{2}Nk_{B}T$, and $(\partial (U/N)/\partial \rho)_{T}=(\partial \phi/\partial \rho)_{T}$. In general the derivative is $T$-dependent but the plateau value, $(\partial \phi_{\infty}/\partial \rho)_{T}$, is a function of density only. Thus (Eq~\ref{Pc}) we propose a density-dependent extended {\it vdW} $a$ coefficient
\begin{eqnarray}
P_{const} = -a(\rho) \rho^{2} \\
 a(\rho) = -(\partial \phi_{\infty}/\partial \rho)
 \label{a0}
 \end{eqnarray}
and the less familiar {\it IS} energy is related to a textbook parameter.

The {\it vdW} repulsive/ideal gas pressure is conventionally derived from the entropy. In the GLH approximation it is easy to see that Eq~\ref{PT} just involves the derivative of the total entropy/particle, $s=s_{c}+s_{v}$, and we propose 
\begin{equation}
P_{T}=-\rho^{2} \frac{\partial}{\partial \rho}(s_{\infty}) \equiv \frac{\rho  k_{B}}{1-\rho b(\rho)},
\label{b}
\end{equation}
with a density-dependent  $b$ coefficient. Since the entropy is evaluated on the plateau, $P_{T}$ is $T$-independent. The contribution to the pressure is $TP_{T}$, with the explicit multiplicative factor coming from the way entropy enters the free energy, $-TS$. It might seem that harmonic oscillations have nothing to do with the molecular congestion that increases the pressure at high density, but the density dependence of $s_{v}$ is entirely due to the change in the shape of the basins, expressed through the frequencies, and that is relevant to packing.

Eq~\ref{b} does not look terribly transparent. For clarification consider that, if 
\begin{equation}
s_{\infty}=k_{B}  ln(\gamma \frac{1-\rho b}{\rho}),
\label{ST}
 \end{equation}
where $\gamma$ is a $\rho$-independent constant, and if for now we let $b$ be $\rho$-independent as well, the {\it vdW} $P_{T}$ is obtained. Eq~\ref{ST} is plausible, with entropy/particle properly varying as $ln(1/\rho)$ at low density and vanishing as ``close packing'' is approached, $\rho \rightarrow 1/b$. Specifically, $s_{\infty}$ vanishes at $\rho=(b+1/\gamma)^{-1}$. However real molecules with soft cores retain positive entropy as they are compressed to very high density,  so any vanishing of $S$ must be understood as an extrapolation from a particular density range. We anticipate that $b(\rho)$ defined in Eq~\ref{b} will decrease at high $\rho$, keeping the entropy positive and the pressure finite. This expectation will be explicitly realized in Sec~\ref{LJ}.

There is considerable interest  \cite{fsmin1,fsmin2,wall} in the variation of the parameter $\alpha$, which determines the total number of {\it IS}, with density and from substance to substance. If the $\rho$-dependence of $\alpha$ (via $\sigma(\phi_{\infty})$) dominates that of $s_{\infty}$,
\begin{equation}
\alpha \approx ln(\gamma \frac{1-\rho b}{\rho}).
\end{equation}

Thus, the free energy evaluated on the high-$T$ plateau of the {\it PEL} yields an equation of state with more complicated density dependence than the {\it vdW} equation, but the same $T$-dependence.  The extended version also takes into account the deviation of $\phi(T)$ from its plateau value, measured by $d\phi(T,\rho) \equiv \phi_{\infty}(\rho)-\phi(T,\rho)$. In the GLH approximation, where $d \phi(T)=\delta^{2}/k_{B}T$, this gives rise to $P_{1/T}\propto (\partial \delta^{2}/ \partial \rho)$. With no density dependent vibrational energy $d \phi(T)$ is the entire relevant energy drop, and it also controls $(S_{c,\infty}-S_{c}(T))$ via $\sigma(\phi_{\infty}-d\phi(T))$; everything is proportional to $\delta^{2}$. Adding the landscape pressure yields the  extended {\it vdW} equation in the GLH approximation,
\begin{equation}
 P(T,\rho)=\frac{k_{B}T\rho}{1-b(\rho)\rho} -a(\rho)\rho^{2}-\rho^{2}\frac{\partial}{\partial \rho}(\frac{\delta^{2}}{2k_{B}T})
 \label{evdw1}
 \end{equation}
 
\medskip
\noindent {\it 2. Beyond the GLH approximation}
\smallskip

Simulation shows that the GLH expression for  $\phi(T)$ is only a crude representation; what is the source of the error? The Gaussian approximation has no upper or lower bound on possible {\it IS} energies, but is a good representation for the contributing fluid-like states over a substantial range of $T$ and $\rho$, and has \cite{derrida,bw,rem} some theoretical justification. 

The harmonic approximation is valid for a particular normal coordinate when the system remains close to the {\it IS}. This suggests high density, where in fact the harmonic $\phi(T)$ is better, and deterioration as $\rho$ is decreased towards the coexistence region. The imaginary frequency instantaneous normal modes \cite{tkinm}, whose number correlates \cite{tkinm,fsh2o,wuli,emh2o1} very well with the self-diffusion coefficient $D$ and increases with decreasing density, are completely anharmonic vibrational coordinates. Thus we aim to retain a Gaussian density of states, but to abandon the linear harmonic approximation for $a_{v}$. Deriving specific  expressions for $a_{v}$ is a current focus of our research but we will now just try to draw some simple, general conclusions.

The true $\phi(T)$ does reach a high-$T$ plateau. Accordingly, we rewrite the free energy,
\begin{eqnarray}
A(T,\rho)/N= [\phi_{\infty} +a_{v}( \phi_{\infty},T,\rho) - k_{B}T\sigma(\phi_{\infty},\rho)] \nonumber \hspace{0in}  \\ - [d\phi(T,\rho)-k_{B}Td\phi^{2}(T,\rho)/2\delta^{2}]-[a_{v}( \phi_{\infty},T,\rho)-  \nonumber \hspace{.2in}  \\ a_{v}( \phi(T,\rho),T,\rho)-k_{B}T (\phi_{0}-\phi_{\infty})d\phi(T,\rho)/\delta^{2}],  \hspace{.4in}
\label{A3}
\end{eqnarray}
suppressing the $\rho$-dependence of the constants $\phi_{\infty}$, $\phi_{0}$ and $\delta^{2}$.  Eq~\ref{A3} has three distinct groups of terms, enclosed by square brackets. The first group is the plateau free energy, which yields $P_{const}$ and $P_{T}$ in the GLH approximation, the second group describes the effect of the drop of $\phi(T)$ from the plateau and becomes the landscape pressure $P_{1/T}$ in the GLH, and the third group vanishes in the GLH.

As a first step consider the form
\begin{eqnarray}
a_{v}(\phi,T,\rho)=(e_{v}^{0}(\rho)+ e_{v}^{1}(\rho) \phi)\nonumber \hspace{.4in} \\ -T(s_{v}^{0}(\rho)+s_{v}^{1}(\rho) \phi)+f(T only);
\label{av}
 \end{eqnarray}
then,
\begin{equation}
\phi(T)=(\phi_{0}+\frac{s_{v}^{1}\delta^{2}}{k_{B}})-\frac{(1+e_{v}^{1}) \delta^{2}}{k_{B}T}.
\label{phiT2}
\end{equation}
Eq~\ref{av} is hardly the most general possibility but it does include linear $\phi$-dependence of both the vibrational energy and entropy, while the linear harmonic approximation has nontrivial  entropy only ($f^{1}=-s^{1}_{v}/k_{B}$, Eq~\ref{phiT1}).  Of course, a function of $T$  only does not influence $P$ or $\phi(T)$. With the Gaussian, Eq~\ref{av} constitutes the Gaussian linear (GL) approximation.
 
Returning to Eq~\ref{A3}, the first group of terms now contains the total (configurational plus vibrational) energy and entropy on the plateau. The vibrational entropy terms in the third group still cancel but the deviation $de_{v}(T)$ of the vibrational energy from the plateau value survives, and combines with $d\phi(T,\rho)$ in the second group to form the deviation of the total energy from the plateau value. The new contribution $de_{v}(T)$ will vanish as the harmonic approximation becomes accurate, i.e, high density. Evaluating the second group with Eq~\ref{phiT2}, the $T$-dependence of $P$ is unchanged from the GLH approximation and
\begin{eqnarray}
P_{const}=\rho^{2}\frac{\partial}{\partial \rho} (\phi_{\infty}+e_{v,\infty}) \label{Pc2} \\
P_{T}=-\rho^{2} \frac{\partial}{\partial \rho}(s_{c,\infty}+s_{v,\infty}) \label{PT2}\\
P_{1/T}=-\rho^{2}\frac{\partial}{\partial \rho}(\frac{(1+e_{v}^{1})^{2}\delta^{2}}{2k_{B}}). \label{P1/T2}
\end{eqnarray}
 The corresponding GL extended {\it vdW} equation is
\begin{equation}
 P(T,\rho)=\frac{k_{B}T\rho}{1-b(\rho)\rho} -a(\rho)\rho^{2}-\rho^{2}\frac{\partial}{\partial \rho}(\frac{(1+e_{v}^{1})^{2}\delta^{2}}{2k_{B}T}).
 \label{evdw2}
 \end{equation}
Eq~\ref{b} still holds for $b$, but
\begin{equation}
a(\rho) = -\frac{\partial}{\partial \rho}(\phi_{\infty}+e_{v,\infty})=-\frac{\partial}{\partial \rho}((1+e_{v}^{1}) \phi_{\infty}+e_{v}^{0}).
\label{a2}
\end{equation}

A more general treatment of anharmonicity is necessary to discuss densities, including the liquid-gas coexistence range, where $\phi(T)$ is not described by  Eq~\ref{phiT2}.  In the GLH and GL approximations there is no characteristic temperature at which the plateau is reached; $d \phi(T)$ vanishes smoothly as $1/T$. Simulation, however, shows \cite{srinat} an identifiable ``top of the landscape'' temperature \cite{caa2}, $T_{ToL}$, at typical liquid densities. Considering the exact Eq~\ref{A3} for the free energy,  if we assume that true plateau quantities are available, they will yield $P_{const}$ and $P_{T}$ according to Eqs~\ref{Pc2} and~\ref{PT2}. It is not terribly difficult to obtain the true $\phi(T)$ from computer simulation, so let us also consider that it is known; then the second group can be evaluated.  The third group, with its cancellations, is less certain. In the GL approximation it converted $d\phi(T)$ to $d\phi(T)+de_{v}(T)$. If we assume that happens more generally we obtain a third Gaussian-anharmonic extended {\it vdW} equation of state
\begin{eqnarray}
 P(T,\rho)=\frac{k_{B}T\rho}{1-b(\rho)\rho} -a(\rho)\rho^{2}- \nonumber \\ \rho^{2} \frac{\partial}{\partial \rho}(d\phi(T,\rho)+de_{v}(T,\rho)-\frac{k_{B}Td\phi^{2}(T,\rho)}{2\delta^{2}}).
 \label{evdw3}
 \end{eqnarray}
The coefficients $a$ and $b$ are given by Eqs~\ref{a2} and~\ref{b}, respectively, with the understanding that the true thermodynamic functions are to be employed. Their properties, and the behavior of $\phi(T)$, implicitly incorporate the effect of an anharmonic $a_{v}$ upon the pressure.

Again, the argument is that, since the GLH approximation for $d\phi(T)$ can be quite poor, the most important step is to correct the contributions to $P$ which directly reflect $d \phi(T)$. On the other hand, contributions evaluated on the plateau are left in the extended {\it vdW} form. They also contain anharmonic effects implicitly through any non-GLH behavior of $\phi_{\infty}(\rho)$ and $s(\phi_{\infty},\rho)$ when they are evaluated, e.g, via simulation.

\section{Some general properties of the extended van der Waals equation}
The extended {\it vdW} equation has  density-dependent $a$ and $b$ coefficients, but its most striking feature is the landscape pressure, in which we now include the $de_{v}(T)$ contribution, with a new $T$-dependence. This term is $\propto 1/T$ in the GLH and GL approximations,  has a more general form in Eq~\ref{evdw3}, and physically expresses the effect on the pressure of the descent of the system from the high-$T$ plateau in the {\it IS} energy. Thus we have a {\it PEL}-based criterion: the extended {\it vdW} equation should provide a good representation of thermodynamic states corresponding to the plateau. In the Angell classification of liquids on a strong-fragile scale \cite{caa2,caa1,fspdb,caa4}, the more complex behavior of fragile liquids is associated with a more pronounced drop $d\phi(T)$ from the plateau. Thus we further suggest that the ordinary vdW equation should be most useful for strong liquids.

\medskip
\noindent {\it 1. Isotherms}
\smallskip

Perhaps the most-discussed features of the {\it vdW} equation are the loops in the isotherms below the critical temperature $T_{c}.$ They are a property of the metastable or unstable homogeneous fluid in the coexistence region. There is no provision in the {\it vdW} equation for the phase separation which goes hand in hand with the true, flat isotherms.

There are two possibilities upon crossing the liquid-gas coexistence curve, $T_{lg}(\rho)$, from above in the ($\rho,T$) plane: the system may phase separate in thermal equilibrium, or it may remain a metastable, homogeneous fluid. Similarly, above the triple point density, cooling below the melting temperature may lead to liquid-solid phase separation or a metastable supercooled liquid.  At a given density, there will exist {\it IS} representative of all the possible thermodynamic states. Under the coexistence curve, there will be both homogeneous fluid and phase-separated liquid-gas and gas-solid {\it IS}. Above the triple point density the {\it IS} types will be homogeneous liquid,  liquid-solid phase separated, and crystal with varying amounts of disorder. Due to surface effects, phase-separated {\it IS} may be difficult to observe, and/or modified in character, in finite-{\it N} simulation. A {\it PEL}-based calculation in a metastable state is achieved by including only the {\it IS} to which the system is restricted, while equilibrium results from keeping all the {\it IS}.

Phase separation causes a large drop in $\phi(T)$, or a strong increase in $d \phi(T)$. On the other hand $d \phi(T)$ varies more gently when the system remains homogeneous. Thus the landscape pressure in Eq~\ref{evdw3} behaves quite differently for the two cases. We suggest that it is small for homogeneous states, making the extended {\it vdW} equation a good approximation, but is large and {\it leads to flat isotherms} if phase separated states are included in the evaluation of $d \phi(T)$.

To demonstrate this idea, let phase separation at a given $\rho$, and the corresponding strong growth in $d \phi(T)$, begin at $T_{lg}(\rho)$. For $T < T_{c}$, the distance to the coexistence curve, $(T_{lg}(\rho)-T)$, is an increasing function of $\rho$ for $\rho < \rho_{c}$ and a decreasing function for $\rho > \rho_{c}$; the density derivative changes from positive to negative at $\rho_{c}$. If  $d \phi(T)$ is a monotonic function of $(T_{lg}(\rho)-T)$, its derivative $(\partial d \phi(T) / \partial \rho)$ will have the same behavior. Then the landscape pressure will be negative below $\rho_{c}$ and positive above, i.e. it will form a {\it van der Waals anti-loop} which can cancel the {\it vdW} loop and produce a flat isotherm. 
\begin{figure}
\psfig{figure=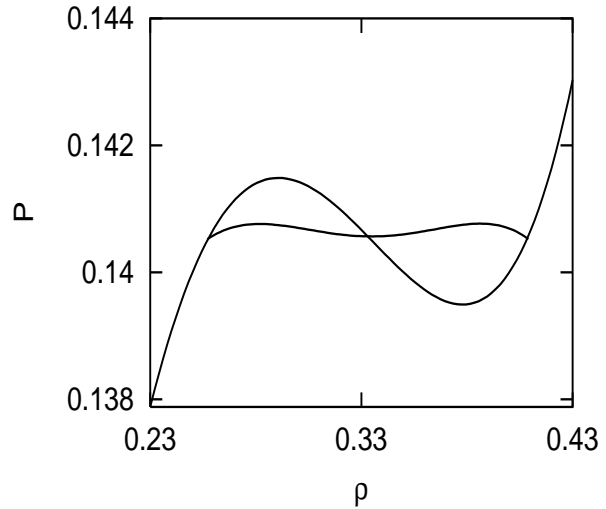,height=2.75in,width=3.25in,angle=-90}
 \caption{Ordinary and extended {\it vdW} pressure $vs$ density for $(T_{c}-T)$=0.015, $a$=4, $b$=1 ($\rho_{c}$=1/3); all quantities in LJ units. Outside the coexistence region the two pressures are identical, inside the flatter curve is the extended {\it vdW} equation with the illustrative form of $d \phi(T,\rho)$ from the text.}
\label{iso}
\end{figure}

As an illustration only, consider the simple anzatz $d \phi(T,\rho)=c(T_{lg}(\rho)-T)^{2}$, with $T_{lg}$ estimated from the ordinary {\it vdW} equation. We use $\rho$-independent $a$=4 and $b$=1 for an approximate description of the LJ fluid and, since the growth of $d \phi(T)$ is the physical effect of interest, ignore $de_{v}$. All quantities are expressed in natural LJ units. The landscape pressure is nonzero within the {\it vdW} coexistence curve only, so we assume it does not change the critical point from $T_{c}=1.185$, $\rho_{c}=0.333$. Fig ~\ref{iso} shows {\it vdW} and extended {\it vdW} isotherms at $T=1.170$ for the choices $c$=1.91, $\delta^{2}$=1. We do not  claim to have the correct $d\phi(T)$ but the point is that any model in which its growth begins at the coexistence curve will yield an anti-loop and potentially a flat isotherm.

\medskip
\noindent {\it 2. Some thermodynamic derivatives}
\smallskip

There is considerable current interest in the phenomonon of a negative $(\partial V/\partial T)_{P}$, the ``density anomaly'' well known in water, which may or may not be associated \cite{anna} with the existence of multiple critical points. Some textbook manipulations show that equivalent conditions are $(\partial S/\partial V)_{T}<0$ or $(\partial P/\partial T)_{V}<0$.

It is immediately apparent that the {\it vdW} equation cannot have a density anomaly, since
\begin{equation}
(\partial P/\partial T)_{V}=\frac{k_{B}\rho}{1-b \rho}>0 \hspace{.15in} (van \ der \ Waals),
\end{equation}
but things are quite different \cite{anomaly} when the landscape pressure is added. In the GL approxination,
\begin{equation}
(\partial P/\partial T)_{V}=\frac{k_{B}\rho}{1-b(\rho)\rho} + \rho^{2}\frac{\partial}{\partial \rho}(\frac{(1+e_{v}^{1})^{2}\delta^{2}}{2k_{B}T^{2}})
\end{equation}
and thus, if the density derivative is negative and large enough, thwre will be an anomaly. A  more complicated condition can be expressed with the true $d \phi(T)$ via Eq~\ref{evdw3}.

There exists a clear physical interpretation of why the extended {\it vdW} equation can have a density anomaly. One expects that entropy should decrease with increasing density (positive $(\partial S/\partial V)_{T}$), as the system becomes more congested, but this is only true on the high-$T$ plateau where $S$ is a maximum. In the GL approximation, the deviation of $\phi(T)$ from the plateau at constant $T$ is proportional to $\delta^{2}$. If $\delta^{2}$ decreases with increasing density, the system gets closer to the maximum plateau entropy, which may compensate for the decrease in $S_{\infty}$ itself, leading to $(\partial S/\partial \rho)_{T}>0$ and $(\partial S/\partial V)_{T}<0$.

For the {\it vdW} equation the energy decreases with increasing density; the relation $(\partial (U/N)/\partial \rho)_{T}=T(\partial P/\partial T)_{V}-P$ yields $(\partial (U/N)/\partial \rho)_{T}=-a$, corresponding to the negative energy of attraction. This is because repulsions enter via the entropy only, i.e, there is no true positive repulsive energy contribution to the pressure, no matter how high the density. In the extended {\it vdW} equation, $(\partial \phi_{\infty}/\partial \rho)$  can (Sec~\ref{LJ}) become large and positive, corresponding to a large negative $a(\rho)$, at high density; the ``{\it vdW} attractive'' term then transforms into a repulsive pressure. Furthermore, taking the landscape pressure into account, in the GL approximation,
\begin{equation}
(\partial (U/N)/\partial \rho)_{T}=-a(\rho)-\frac{\partial}{\partial \rho}(\frac{(1+e_{v}^{1})^{2}\delta^{2}}{k_{B}T}),
\end{equation}
which can possibly change sign for the same reason given in the last paragraph for the density anomaly.

The critical behavior of the heat capacity  is described \cite{hesbook} by the exponents $\alpha$ and $\alpha'$,
\begin{eqnarray}
(\partial U/\partial T)_{V_{c}} \propto (T-T_{c})^{-\alpha}, \ T>T_{c} \nonumber \\
\propto (T_{c}-T)^{-\alpha'}, \ T<T_{c} \nonumber
\end{eqnarray}
In the {\it vdW} approximation \cite{hesbook}, $\alpha$=$\alpha'$= 0. Since we are obtaining $P$ from the free energy the simplest consistent route to $U$ is also through $A$; elementary statistical mechanics yields the exact relation
\begin{equation}
(U(T)/N) = \phi(T) - T^{2} (\frac{\partial (a_{v}(T,\phi)/T)}{\partial T})_{\phi(T)}
\label{U}
\end{equation}
where the notation is intended to convey that the derivative acts on the explicit $T$-dependence of $a_{v}$ only, not on that which enters implicitly through $\phi(T)$. For an equation of state based upon a particular approximation to $a_{v}$, use of that approximation in Eq~\ref{U} yields the corresponding potential energy. Since the {\it vdW} equation is obtained by ignoring the deviations of $\phi$ and $e_{v}$ from their plateau values in $a_{v}$,  strong $T$-dependence of $U(T)$ is discarded. That is why $\alpha$=$\alpha'$= 0; the critical anomaly is in the landscape term.

\section{Application to the Lennard-Jones liquid \label{LJ}}

\begin{figure}
\psfig{figure=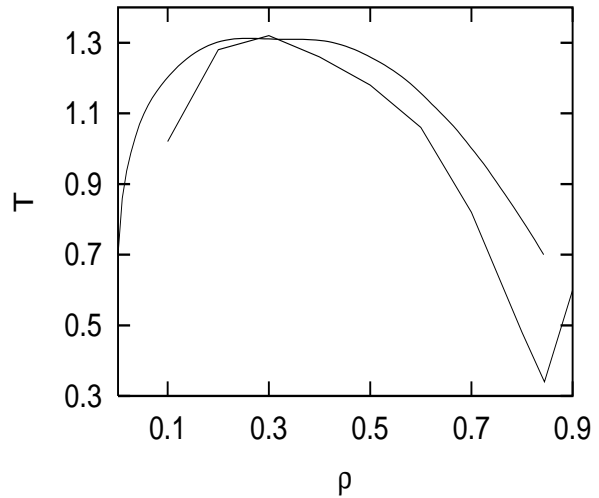,height=2.75in,width=3.25in,angle=-90}
 \caption{Simulated $T_{ToL}(\rho)$ data (jagged) and coexistence curve from Ref ~\cite{LJeos}, LJ units.}
\label{tol}
\end{figure}
Here we present some preliminary results on LJ. A careful study of phase equilibria  \cite{LJeos} determined that the critical point and triple point are $\rho_{c}$=0.31, $T_{c}$=1.31 and $\rho_{t}$=0.84, $T_{t}$=0.75,  LJ units.  An estimate of $T_{ToL}(\rho)$, determined from $\phi(T)$, {\it N=256}, is shown in Fig~\ref{tol}, along with the coexistence curve from Ref~\cite{LJeos}. A minimum is found near the triple point density, and with decreasing density $T_{ToL}(\rho)$ runs between the liquid-gas spinodal and coexistence curve to the critical point.  Earlier we discussed the liquid-gas coexistence temperature, $T_{lg}$, as marking the onset of a rapid rise in $d \phi(T)$; $T_{lg}$ is a ``{\it ToL}'' temperature for {\it equilibrium} states. Practically, the metastable homogeneous phase can survive until the spinodal is reached, and our result is expected in finite-size simulation. What may be interesting is that the liquid-gas spinodal and the {\it ToL} temperature discussed in supercooled liquids, found at densities above the minimum, are thus connected.  Leyvraz and Klein \cite{billk} have suggested that properties of supercooled liquids may be influenced by a spinodal. 

\begin{figure}
\psfig{figure=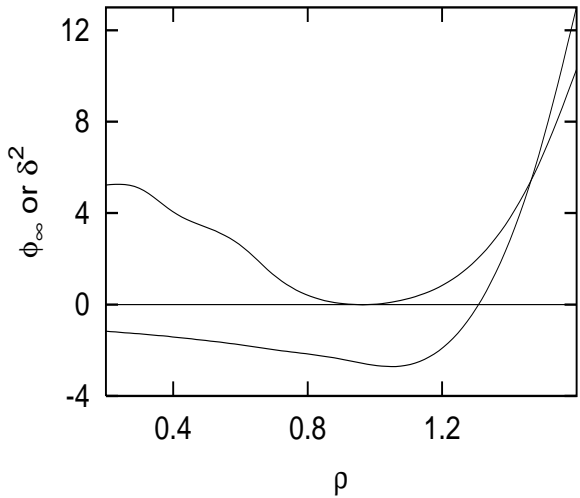,height=2.75in,width=3.25in,angle=-90}
 \caption{Plateau energy $\phi_{\infty}$ (lower), shifted up by 5 energy units, and squared Gaussian width $\delta^{2}$ {\it vs} reduced density, all in LJ units}
\label{del2}
\end{figure}

We have obtained $\delta^{2}(\rho)$ and $\phi_{\infty}(\rho)$  (Fig~\ref{del2}) for 2.0$\ge$$\rho$$\ge$0.20, {\it N=500}, from the distribution of {\it IS} visited at high $T$, where $\delta^{2}$ becomes $T$-independent, and from $\phi(T)$, respectively.  Starting at low density both quantities, like $T_{ToL}$, decrease to minmium values near the triple point, and then begin to rise. Similar behavior has been observed \cite{pdbeos,Pismin} for the {\it IS} pressure. Correspondingly the {\it vdW} attractive term becomes a positive, repulsive pressure. In the GLH approximation the behavior of $\delta^{2}(\rho)$ would cause the landscape pressure to vanish near the triple point, identifying a region well described by the extended {\it vdW} equation.

The higher the density, the better we expect the harmonic approximation to perform. Thus we are going to use our data to evaluate the GLH equation of state over the entire available density range, but but we do not make any claims of validity at low to intermediate density. Fitting $\delta^{2}(\rho)$ and $\phi_{\infty}(\rho)$ to a sum of two exponentials gives good fits, particularly at high density, 2.0$\ge$$\rho$$\ge$1.2.  Taking density derivatives yields $P_{const}$ ($a(\rho)$) and the landscape pressure $P_{1/T}$.

We do not at the moment have the information necessary to calculate $b(\rho)$ from first principles. This coefficient involves an extrapolation, i.e, in some small density range it appears that the pressure would diverge at a particular, higher ``close packed'' density, but since the cores are soft that density is never reached. Consequently $b$ may appear to be constant at low density, but its $\rho$-dependence becomes essential at high density. 

\begin{figure}
\psfig{figure=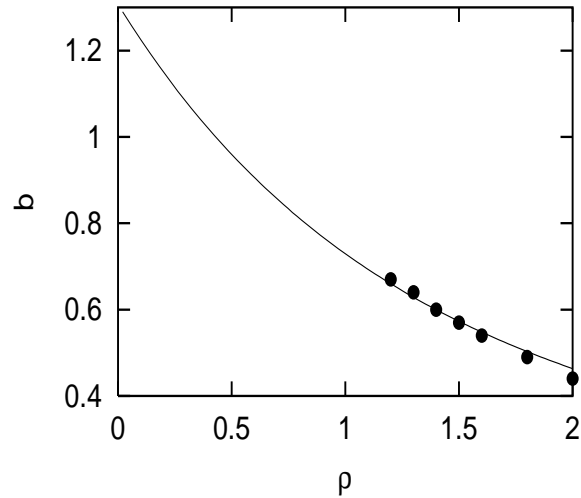,height=2.75in,width=3.25in,angle=-90}
 \caption{High density $b$ coefficient from simulation and fit to Eq~\ref{b1rho}, $b_{0}$=1.31. Note $b$($\rho_{c}$=0.31)=1.08.}
\label{brho}
\end{figure}

It is difficult to equilibrate the system at $\rho$$>$1.2 without going to quite high $T$. To get some idea of the behavior of $b$ we have calculated the $T$=25 isotherm, and $b(\rho)$ from the GLH Eq~\ref{evdw1}. The results are shown in Fig~\ref{brho}. Because of the soft cores, the quantity $(1-b(\rho)\rho)$ can become small, but never zero. Thus we suggest a plausible behavior is exponential decay with $\rho$, giving
\begin{equation}
b(\rho)=(1-e^{-b_{0}\rho})/ \rho,
\label{b1rho}
\end{equation}
and the smooth curve in Fig~\ref{brho} is Eq~\ref{b1rho} with $b_{0}$=1.31. Combining everything,
\begin{eqnarray}
P(T,\rho)=T \rho e^{1.31 \rho} +  \rho^{2}(9.15e^{1.95 \rho}-13.7e^{1.48 \rho}) \nonumber \\
- \frac{ \rho^{2}}{T} (0.0168e^{4.96 \rho}-35.4e^{-3.09}), \hspace{.5in} 
\label{hiro}
\end{eqnarray}
anticipating that the analog of the {\it vdW} attractive term will be positive, and the landscape term negative.

\begin{figure}
\psfig{figure=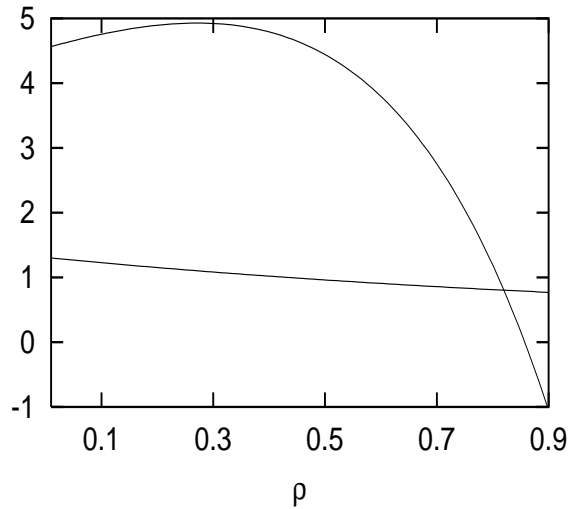,height=2.75in,width=3.25in,angle=-90}
 \caption{$a$ (upper) and $b$ coefficients {\it vs} density}
\label{abrho}
\end{figure}

Matching the \cite{LJeos} true $\rho_{c}$ and $T_{c}$ of LJ to the {\it vdW} equation gives $b$=1.07 and $a$=4.73 At the critical density we find $a(0.31)$=4.92 (negative of first parenthesis in Eq~\ref{hiro}) and (Eq~\ref{b1rho})  $b(\rho)$=1.08. Again, we expect the missing anharmonicity to be important at the critical density and regard these results with some skepticism, but the agreement is remarkable. The density dependence of $a$, along with $b$, is shown for $\rho \le 0.90$ in Fig~\ref{abrho}; note the sign change near the triple point density.

\section{summary}

The equation of state is obtained from the density derivative of the Helmholtz free energy, as always. The {\it PEL} approach simply provides a less traditional way to view the problem and generate approximations. Since a Gaussian approximation for the {\it IS} energy distribution is reasonable, the focus is on the vibrational free energy $a_{v}$. The GLH approximation was worked out \cite{emeos} by La Nave et al., and  the starting point of this paper is simply the observation that  their result resembles the van der Waals equation with density-dependent $a$ and $b$ coefficients, and an extra ``landscape'' term.

In addition, we have tried to include the anharmonicity of $a_{v}$ with as few assumptions as is possible; in the extended {\it vdW} Eq~\ref{evdw3}, anharmonicity is implicit in the $T$-dependent inherent structure energy, and in the plateau entropy. The smaller the drop of the {\it IS} energy from its high-$T$ plateau, the smaller the landscape pressure, suggesting that the {\it vdW} equation is best suited to strong liquids. The extended  equation can reproduce, and provide {\it PEL} interpretations of, thermodynamic phenomena absent from the usual {\it vdW} equation.

Computer simulation in the LJ fluid yields $a(\rho)$ and $b(\rho)$. The $a$ coefficient becomes negative at $\rho \gtrsim  1.0$, as the ``{\it vdW} attractive'' pressure becomes a repulsive pressure. The $b$ coefficient is represented by a simple expression stemming from the idea that the pressure may become exponentially large, but not infinite. Since the harmonic approximation may be accurate at high density, the resulting analytic expression is suggested as a high density equation of state, which we will explore in future work.  Nevertheless, when evaluated at the critical density, $a(\rho_{c})$ and $b(\rho_{c})$ are remarkably close to the accepted values.

The quantities $\phi_{\infty}(\rho)$ and $\delta^{2}(\rho)$ have minima near the triple point. Thus in the GLH approximation both $P_{const}$ ($a$) and $P_{1/T}$ (landscape pressure) vanish. The ``top of the landscape'' temperature also has a minimum near the triple point. At lower density $T_{ToL}(\rho)$ runs under the coexistence curve to the critical point, at higher density it rises and becomes the quantity discussed \cite{caa2} for supercooled liquids; this connection may support the conjecture \cite{billk} of Klein and Leyvraz.

\medskip
\noindent {\bf Acknowledgment}
\smallskip

This work was supported by NSF Grant CHE-0090975. Discussions with F. Sciortino are greatfully acknowledged.

\end{document}